\documentclass[10pt]{article}

\usepackage{authblk}
\usepackage{natbib}
\usepackage{mathtools}
\usepackage{amsmath}
\usepackage{amsthm}
\usepackage{amssymb}
\usepackage{amsbsy}
\usepackage{amsfonts}
\usepackage{amscd}
\usepackage{mathrsfs}
\usepackage{bm}
\usepackage{breqn}
\usepackage{color, soul}
\usepackage{enumerate}
\usepackage{empheq}
\usepackage{rotating}
\usepackage{ftnxtra}
\usepackage{fnpos}
\usepackage[titletoc,toc,title]{appendix}
\usepackage{euscript}
\usepackage{graphicx}
\usepackage{epsfig}
\usepackage{epstopdf}
\DeclareGraphicsExtensions{.pdf,.png,.jpg,.eps}
\usepackage{pstool}
\usepackage{upgreek}
\usepackage{mathrsfs}

\usepackage{psfrag}

\numberwithin{equation}{section}

\theoremstyle{plain}	
\newtheorem{thm}{Theorem}[section]

\newtheorem{prop}[thm]{Proposition}
\newtheorem*{prop*}{Proposition} 
\theoremstyle{definition}	
\newtheorem{definition}[thm]{Definition}
\newtheorem{remark}[thm]{Remark}
\newtheorem{example}[thm]{Example}

\usepackage{graphicx}
\usepackage{lipsum}

\usepackage{caption}
\usepackage{subcaption}

\usepackage{hyperref}
\hypersetup{colorlinks=true, linkcolor=blue}
\hypersetup{colorlinks=true,citecolor=blue}

\DeclareMathAlphabet{\mathpzc}{OT1}{pzc}{m}{it}

\usepackage{amsmath, amsthm, amssymb}

\usepackage{cleveref}

\DeclarePairedDelimiter\abs{\lvert}{\rvert}

\makeatletter
\newsavebox{\@brx}
\newcommand{\llangle}[1][]{\savebox{\@brx}{\(\m@th{#1\langle}\)}%
  \mathopen{\copy\@brx\mkern2mu\kern-0.9\wd\@brx\usebox{\@brx}}}
\newcommand{\rrangle}[1][]{\savebox{\@brx}{\(\m@th{#1\rangle}\)}%
  \mathclose{\copy\@brx\mkern2mu\kern-0.9\wd\@brx\usebox{\@brx}}}%
\let\oldabs\abs
\def\abs{\@ifstar{\oldabs}{\oldabs*}}
\makeatother

\usepackage{accents}

    {\end{bmatrix}}%

\usepackage{enumitem}

\usepackage[utf8]{inputenc}
\usepackage{dutchcal}
\usepackage{multirow}

\usepackage{xcolor}

\usepackage{setspace}

\begin{document}

\title{\textbf{The Darboux Classification of Curl Forces}}

\author[1]{Arash Yavari}
\author[2]{Alain Goriely\thanks{Corresponding author, e-mail: alain.goriely@maths.ox.ac.uk}}
\affil[1]{\small \textit{School of Civil and Environmental Engineering and The George W. Woodruff School of Mechanical Engineering, Georgia Institute of Technology, Atlanta, GA 30332, USA,}}\affil[2]{\small \textit{Mathematical Institute, University of Oxford, Oxford, OX2 6GG, UK}}

\maketitle

\begin{abstract}
We study particle dynamics under curl forces. These forces are a class of non-conservative, non-dissipative, position-dependent forces  that cannot be expressed as gradient of a potential function.   We show that the fundamental quantity of particle dynamics under curl forces is a \textit{work $1$-form}.
By using the Darboux classification of differential $1$-forms on $\mathbb{R}^2$ and $\mathbb{R}^3$, we establish that any curl force in two dimensions has at most two \textit{generalized potentials}, while in three dimensions, it has at most three. These potentials generalize the single potential of conservative systems.  For any curl force field, we introduce a corresponding conservative force field---\textit{the conservative auxiliary force}. 
The Hamiltonian of this conservative force is a conserved quantity of motion for the dynamics of a particle under the curl force, although it is not the physical energy.
\end{abstract}

\begin{description}
\item[Keywords:] Nonconservative force, curl force, Darboux classification, generalized potential energies.
\end{description}


\section{Introduction}

In classical mechanics, a conservative force field is one that can be derived from a potential energy. Typical examples of non-conservative forces, such as friction and viscous forces, are dissipative. However, a non-conservative force---one that does not have an associated potential---is not necessarily dissipative. An important class of such non-conservative, non-dissipative force fields are forces of the form $\mathbf{F} = \mathbf{F}(\mathbf{x})$, depend on position $\mathbf{x} \in \mathbb{R}^n$ ($n=2, 3$) but not on velocity, and for which $\operatorname{curl}\, \mathbf{F}(\mathbf{x}) \neq \mathbf{0}$. Following \cite{Berry2012}, we call these force fields \emph{curl forces}. 
Non-conservative forces that depend on position are also referred to as \textit{positional forces}, \textit{pseudo-gyroscopic forces}, and \textit{circulatory forces} \citep{Ziegler1977,Kirillov2021}.
These forces are non-conservative yet non-dissipative, meaning that they preserve the phase space volume. Another characterization of non-conservative, non-dissipative forces is provided in \S\ref{Sec:Work}.
Particle dynamics under such force fields differs fundamentally from that under conservative forces. For instance, Noether’s theorem, which connects conservation laws with symmetries, cannot be used in this context \citep{Berry2012, Berry2013}. 
Curl forces in both discrete and continuous mechanical systems such as Cauchy elasticity have a rich history. For a recent detailed literature review, see \citep{yavari2024nonlinear}.

For a conservative force field $\mathbf{F}=\mathbf{F}(\mathbf{x})$, there exists a potential energy $U=U(\mathbf{x})$ such that
\begin{equation}
    \mathbf{F}(\mathbf{x})=-\nabla U=-\frac{\partial U}{\partial\mathbf{x}}
    \,.
\end{equation}
For a curl force, no such potential energy  exists. However, we will show in the following that any curl force has at most two and three \emph{generalized potential energies} in $2D$ and $3D$, respectively.

\citet{Berry2015} showed that a subset of curl forces admits Hamiltonians of the form of an anisotropic kinetic energy + a scalar potential. 
In this paper, we demonstrate that any curl force can be written in terms of three generalized potentials and admits an auxiliary Hamiltonian which, although not representing the physical energy, is nevertheless a conserved quantity of motion (albeit an implicit one that does not divide the phase-space into invariant regions).

This paper is organized as follows. In \S\ref{Sec:Classification}, we present the Darboux classification of curl forces in two and three dimensions, establishing the fundamental structure of the work $1$-form. In \S\ref{Sec:Darboux-2D} and \S\ref{Sec:Darboux-3D}, we calculate the generalized potentials associated with two- and three-dimensional curl forces, respectively. 
In \S\ref{Sec:Work}, we study the work performed by curl forces in closed and cyclic motions. In \S\ref{Sec:KineticEnergy}, we investigate the change in the kinetic energy of a particle under curl forces and interpret it in the context of Carath\'eodory's formulation of thermodynamics. 
The local accessibility property of curl forces is discussed in \S\ref{Sec:Accessibility}. Finally, \S\ref{Sec:Hamiltonian} introduces a generalized Hamiltonian formulation for curl forces, where we construct an auxiliary conservative force field and derive the corresponding Hamiltonian. 
We first analyze the two-dimensional case in \S\ref{Sec:Hamiltonian2D}, including the calculation of generalized potentials, before extending the formulation to three-dimensional curl forces in \S\ref{Sec:Hamiltonian3D}. Some concluding remarks are given in \S\ref{Sec:Conclusions}.

\section{Canonical forms of curl forces in two and three dimensions} \label{Sec:Classification}

Let $\boldsymbol{\alpha}$ be a differential form on an $n$-manifold $\mathcal{M}$. We define
\begin{equation}
    (\boldsymbol{\alpha})^{k}=\overbrace{\boldsymbol{\alpha} \wedge \hdots \wedge \boldsymbol{\alpha}}^{k-\text{factors}}
    \,.
\end{equation}
Suppose $\boldsymbol{\Omega}$ is a $1$-form on $\mathcal{M}$. $k$ is called the rank of $\boldsymbol{\Omega}$ if $(d\boldsymbol{\Omega})^k\neq 0$ and $(d\boldsymbol{\Omega})^{k+1}= 0$. 
Note that $(d\boldsymbol{\Omega})^k$ is a $2k$-form and the condition $(d\boldsymbol{\Omega})^k\neq 0$ implies that $n>2k$ or $n-k>k$.

\begin{thm}[Darboux \citep{Darboux1882,slebodzinski1970exterior,Sternberg1999,Bryant2013,Suhubi2013}]
Let $\boldsymbol{\Omega}$ be a $1$-form of rank $k$ on an $n$-manifold $\mathcal{M}$. Suppose $\boldsymbol{\Omega}\wedge(d\boldsymbol{\Omega})^k= 0$ everywhere on $\mathcal{M}$. Then in a neighbourhood of any point, there exist coordinates $\{y^1,\hdots,y^{n-k},z^1,\hdots,z^k\}$ such that
\begin{equation} \label{Canonical1}
    \boldsymbol{\Omega}=y^1dz^1+\hdots+y^k dz^k
    \,.
\end{equation}
If $\boldsymbol{\Omega}\wedge(d\boldsymbol{\Omega})^k\neq 0$ everywhere on $\mathcal{M}$, then in a neighbourhood of every point there exist coordinates $\{y^1,\hdots,y^{n-k},$ $z^1,\hdots,z^k\}$ such that
\begin{equation} \label{Canonical2}
    \boldsymbol{\Omega}=y^1dz^1+\hdots+y^kdz^k+dy^{k+1}
    \,.
\end{equation}
\end{thm}

Eqs.~\eqref{Canonical1} and \eqref{Canonical2} are referred to as the canonical (or normal) forms of $\boldsymbol{\Omega}$.

\begin{definition}[Work $1$-form]
Given a force field $\mathbf{F}=\mathbf{F}(\mathbf{x})$ in $\mathbb{R}^n$ ($n=2$ or $3$), the work $1$-form is defined as $\boldsymbol{\Omega}=\mathbf{F}(\mathbf{x}) \cdot d\mathbf{x}$. Explicitly, in a (curvilinear) coordinate chart $\{x^a\}$ with metric components $g_{ab}=g_{ba}$, the work 1-form has the representation $\boldsymbol{\Omega}=F^b g_{ba}\,dx^a= F_a \,dx^a$. The work 1-form can also be written in a coordinate-free representation as $\boldsymbol{\Omega}=\mathbf{F}^\flat$, where the flat operator $\flat$  is an isomorphism between vectors and $1$-forms.  \end{definition}

We note that the analog of the work $1$-form in Cauchy elasticity is the stress work $1$-form \citep{yavari2024nonlinear}.

A force field $\mathbf{F}=\mathbf{F}(\mathbf{x})$ in $\mathbb{R}^n$ is a curl force field if and only if $d\boldsymbol{\Omega}\neq\mathbf{0}$, where $d$ is the exterior derivative and $d\boldsymbol{\Omega}$ is a $2$-form. With respect to a coordinate chart $\{x^a\}$, $d\boldsymbol{\Omega}=F_{a,b} \,dx^b \wedge dx^a$, where $\wedge$ is the wedge product of differential forms. 
In Euclidean space, the curl of a vector field $\mathbf{F}=F^a\frac{\partial}{\partial x^a}$ has the following relation with exterior derivative \citep{Abraham2012}
\begin{equation} \label{Curl-F}
	\operatorname{curl}\mathbf{F} = \left[ \star(d \mathbf{F}^\flat) \right]^\sharp\,,
\end{equation}
where $\sharp$ is the sharp operator, which is an isomorphism between $1$-forms and vectors (the inverse of the $\flat$ operator), and $\star$ is the Hodge star operator, which is an isomorphism between $k$-forms and $(n-k)$-forms ($n=2,3$) and for any $1$-form $\boldsymbol{\alpha}$, $\star\star\boldsymbol{\alpha}=(-1)^{n-1}\boldsymbol{\alpha}$.
Eq.~\eqref{Curl-F} can be rewritten as
\begin{equation} 
	d \mathbf{F}^\flat = d\boldsymbol{\Omega} = (-1)^{n-1}  
	\star\left[(\operatorname{curl}\mathbf{F})^\flat\right]    
	\,.
\end{equation}
Therefore, $d\boldsymbol{\Omega} \neq \mathbf{0}$ if and only if $\operatorname{curl}\mathbf{F}\neq \mathbf{0}$.

Curl force fields are classified for $n=2$ and $n=3$ using Darboux's theorem as follows:

\begin{prop}
In  $\mathbb{R}^2$, any  force field $\mathbf{F}= \mathbf{F}(\mathbf{x})$  has the following representation:
\begin{equation} \label{Curl-Force-2D}
    \mathbf{F}(\mathbf{x})=-V(\mathbf{x})\,\nabla U(\mathbf{x})        \,,
\end{equation}
where $U(\mathbf{x})$ and $V(\mathbf{x})$ are  the \textit{generalized potentials}, with $\nabla V(\mathbf{x})=\mathbf 0$ if an only if $\operatorname{curl}\mathbf{F}=\mathbf 0$.
More generally, $\operatorname{curl} \mathbf{F} = \mathbf{0}$ if an only if $V=f(U)$ for some function $f$.
\end{prop}

\begin{proof}
If $\operatorname{curl}\mathbf{F}=\mathbf 0$, then $\mathbf{F}$ can be written as $\mathbf{F}(\mathbf{x})=-\nabla U(\mathbf{x}) $ which implies that $V=1$ and $\nabla V=\mathbf{0}$.
If $\operatorname{curl}\mathbf{F}\not=\mathbf 0$, then in  $2$D, $d\boldsymbol{\Omega}\neq\mathbf{0}$ but $d\boldsymbol{\Omega}\wedge d\boldsymbol{\Omega}$ is a $4$-form, which identically vanishes on $\mathbb{R}^2$. Thus, $\boldsymbol{\Omega}$ has rank $k=1$. Also, note that $\boldsymbol{\Omega}\wedge d\boldsymbol{\Omega}$ is a $3$-form, which vanishes on $\mathbb{R}^2$. Therefore, $\boldsymbol{\Omega}$ has the canonical form $\boldsymbol{\Omega}(\mathbf{x})=\phi(\mathbf{x}) \,d\psi(\mathbf{x})$. 
Thus, $\mathbf{F}^\flat=\phi(\mathbf{x}) \,d\psi(\mathbf{x})$, and hence $\mathbf{F}=\phi(\mathbf{x}) \,[d\psi(\mathbf{x})]^\sharp$. Recalling that $[d\psi(\mathbf{x})]^\sharp=\nabla\psi(\mathbf{x})$, one concludes that any curl force field in $\mathbb{R}^2$ has the representation \eqref{Curl-Force-2D}, where $V(\mathbf{x})=-\phi(\mathbf{x})$ and $U(\mathbf{x})=\psi(\mathbf{x})$. 
Note that\footnote{It is known that for arbitrary vectors $\mathbf{u}$ and $\mathbf{w}$ one has $\mathbf{u}\times\mathbf{w}=[\star(\mathbf{u}^\flat\wedge \mathbf{w}^\flat)]^\sharp$ \citep{Abraham2012}.}
\begin{equation}
	\nabla V \times \nabla U 
	= \left[\star\left((\nabla V)^\flat\wedge (\nabla U)^\flat \right) \right]^\sharp
	= \left[\star\left(-d\phi \wedge d\psi \right) \right]^\sharp
	=-\left[\star \left(d\boldsymbol{\Omega} \right) \right]^\sharp \,.
\end{equation}
This implies that $d\boldsymbol{\Omega}=d\phi\wedge d\psi\neq\mathbf{0}$ if and only if $\nabla V \times \nabla U\neq \mathbf{0}$. 
It is clear that if $\nabla V(\mathbf{x}) = \mathbf{0}$, then $d\boldsymbol{\Omega} = \mathbf{0}$, and hence $\operatorname{curl} \mathbf{F} = \mathbf{0}$.
Note that $d\phi \wedge d\psi = \mathbf{0}$ if and only if $\phi = \phi(\psi)$, i.e., $\phi$ and $\psi$ are functionally dependent. This, in turn, implies that $\operatorname{curl} \mathbf{F} = \mathbf{0}$ if and only if $V = f(U)$ for some function $f$. In other words, \eqref{Curl-Force-2D} represents a curl force if and only if the two potentials are functionally independent.
\end{proof}

\begin{prop} 
In $\mathbb{R}^3$, any force field $\mathbf{F}= \mathbf{F}(\mathbf{x})$ has the following representation:
\begin{equation} \label{Curl-Force-3D}
    \mathbf{F}(\mathbf{x})= -V(\mathbf{x})\,\nabla U(\mathbf{x})-\nabla W(\mathbf{x})    \,,
\end{equation}
where $U(\mathbf{x})$, $V(\mathbf{x})$ and $W(\mathbf{x})$, provided that $\nabla V \times \nabla U\neq \mathbf{0}$, are called the generalized potentials. In particular,   $\nabla V(\mathbf{x})=\mathbf 0$ if and only if $\operatorname{curl}\mathbf{F}=\mathbf 0$ (in which case we can take $W=0$ without loss of generality), and  $\nabla W(\mathbf{x})=\mathbf 0$  if and only if   $\mathbf{F}\cdot\operatorname{curl}\mathbf{F}=0$.
\end{prop}

\begin{proof}
If $\operatorname{curl}\mathbf{F}=\mathbf 0$, then $\mathbf{F}$ can be written as  $\mathbf{F}(\mathbf{x})=-\nabla U(\mathbf{x}) $ which implies that $V=1$ and $W=0$.
For a $3$D curl force, $d\boldsymbol{\Omega}\neq\mathbf{0}$ but $d\boldsymbol{\Omega}\wedge d\boldsymbol{\Omega}$ is a $4$-form, which vanishes on $\mathbb{R}^3$. Thus, again $\boldsymbol{\Omega}$ has rank $k=1$. We have the following two possibilities 
\begin{equation}
    \begin{dcases}
    \boldsymbol{\Omega}\wedge d\boldsymbol{\Omega} =0    
    ~ \Rightarrow ~ \boldsymbol{\Omega}=\phi\, d\psi\,,\\
    \boldsymbol{\Omega}\wedge d\boldsymbol{\Omega} \neq 0   
    ~ \Rightarrow ~ \boldsymbol{\Omega}=\phi\, d\psi+d\zeta \,.
    \end{dcases}
\end{equation}
Therefore, the force field has the representation \eqref{Curl-Force-3D}, where $V=-\phi$, $U=\psi$, and $W=-\zeta$.
Note that\footnote{One can show that for arbitrary vectors $\mathbf{u}$ and $\mathbf{w}$ one has $(\mathbf{u}\cdot\mathbf{w}) \boldsymbol{\mu}=\mathbf{u}^\flat \wedge \star(\mathbf{w}^\flat)$, where $\boldsymbol{\mu}$ is the volume element of the Euclidean space ($\boldsymbol{\mu}=dx\wedge dy$ and $\boldsymbol{\mu}=dx\wedge dy \wedge dz$ in $2$D and $3D$, respectively) \citep{Abraham2012}.}
\begin{equation}
	\left(\mathbf{F}\cdot\operatorname{curl}\mathbf{F}\right)\boldsymbol{\mu}
	= \mathbf{F}^\flat \wedge \star \left(\operatorname{curl}\mathbf{F}\right)^\flat
	= \boldsymbol{\Omega}\wedge \star \left[ \star(d \mathbf{F}^\flat) \right]
	= (-1)^{n-1}  \boldsymbol{\Omega}\wedge d\boldsymbol{\Omega}\,.
\end{equation}
This implies that $\boldsymbol{\Omega}\wedge d\boldsymbol{\Omega} =0$ is equivalent to $\mathbf{F}\cdot\operatorname{curl}\mathbf{F}=0$.
\end{proof}

\begin{remark}
We note that the quantity $\chi(\mathbf F)= \mathbf{F}\cdot\operatorname{curl}\mathbf{F}$ that determines whether or not a force derives from one ($V=1,W=0$), two ($W=0$), or three potentials, is referred to as the helicity or chirality of the field $\mathbf F$ and plays a central role in fluid mechanics \citep{Moffatt1969,MoffattTsinober1992,ArnoldKhesin1998} and electromagnetism \citep{TruebaRanada1996}. 
If the force field is chiral $\chi(\mathbf F)\not= 0$, then three potentials are needed to represent it. These three functions can be thought of as an alternative representation of the three components of $\mathbf F$ that capture the work performed during motion.
\end{remark}

\subsection{Calculation of the generalized potentials for $2$D curl forces} \label{Sec:Darboux-2D}


From $\mathbf{F}=-V\,\nabla U$, we have $\nabla U= - \frac{\mathbf{F}}{V}$. For this PDE to have a solution the following integrability (compatibility) conditions must be satisfied:\footnote{We are tacitly assuming that the domain is simply connected (in particular, contractible), so that  Poincar\'e's lemma applies and a closed $1$-form is exact. However, if one works instead on a non-simply connected subset $\mathcal{U} \subset \mathbb{R}^n$, this is \emph{not} generally sufficient.  In that case, even when $\nabla \times \left(\frac{\mathbf{F}}{V}\right)=\mathbf{0}$, one must also check that the period (integral) of the $1$-form $\boldsymbol{\alpha} = \tfrac{\mathbf{F}}{V}\cdot d\mathbf{x}$ around every non-contractible closed loop in $\mathcal{U}$ vanishes. Equivalently, this means that the cohomology class $[\boldsymbol{\alpha}]$ in the first de Rham cohomology group $H^1(\mathcal{U})$ must be zero. Only when both conditions hold---curl-free and all periods vanish---does there exist a globally defined scalar function $U$ satisfying $\nabla U = -\tfrac{\mathbf{F}}{V}$ \citep{CantarellaDeturckGluck2002,Lee2013,Yavari2013}.}
\begin{equation}
	\nabla \times \left( \frac{\mathbf{F}}{V} \right) 
	= \frac{1}{V} (\nabla \times \mathbf{F}) - \frac{\nabla V}{V^2} \times \mathbf{F}=\mathbf{0}
	\,.
\end{equation}
Therefore, the generalized potential $V$ must satisfy the following first-order PDE:
\begin{equation} \label{V-PDE}
	\nabla V \times \mathbf{F} - V \,\nabla \times \mathbf{F} =\mathbf{0}
	\,.
\end{equation}
Solving the above PDE gives $V$, after which $\nabla U= - \frac{\mathbf{F}}{V}$.

\begin{example}\label{Force-Field-Calculation}
Let us consider the following curl force field:\footnote{\citet{Berry2015} considered a similar example as a force field that cannot be treated using their anisotropic kinetic energy approach.}
\begin{equation} \label{Curl-Force-Berry}
	\mathbf{F}(x,y)=(F_x,F_y) = -\frac{F_0}{a^3} (xy^2,x^3)	\,,
\end{equation}
where $F_0$ and $a$ are positive constants with physical dimensions of force and length, respectively. 
The curl of this force field is $-\frac{F_0}{a^3}(3 x^2 - 2 x y)$ ($z$-component of curl).
The PDE \eqref{V-PDE} for this force field is simplified to read
\begin{equation}
	x^2 \frac{\partial V}{\partial x} - y^2 \frac{\partial V}{\partial y} - (3x - 2y) V = 0\,.
\end{equation}
Let us rewrite this PDE in the form of a characteristic equation \citep{Duff1956}
\begin{equation}
	\frac{dx}{x^2} = \frac{dy}{-y^2} = \frac{dV}{(3x - 2y) V}
	\,.
\end{equation}
The first equality gives us the characteristic curves
\begin{equation}
	\frac{x+y}{xy} =c
	\,.
\end{equation}
The solution of the PDE is
\begin{equation}
	V(x,y)=x^3y^2\,\Phi\!\left(\frac{x+y}{xy} \right)
	\,,
\end{equation}
where $\Phi$ is an arbitrary differentiable function. 
Thus,
\begin{equation}
	\nabla U(x,y) 
	= \frac{F_0}{a^3\, \Phi\!\left(\frac{x+y}{xy} \right)} \left(\frac{1}{x^2},\frac{1}{y^2} \right)
	\,.
\end{equation}
We see that $U$ is not unique. This is not surprising as $V\,\nabla U$ is invariant under the transformations $(U,V)\mapsto \left(f(U),\frac{V}{f'(U)}\right)$ for any differentiable function $f$ such that $f'(U)\neq 0$.
A choice would be $\Phi=a^{-5}$, i.e., $V(x,y)= a^{-5} x^3y^2$. This implies that $\nabla U= F_0a^2(\frac{1}{x^2},\frac{1}{y^2})$, and hence 
\begin{equation} \label{U-Example}
	U(x,y)= -F_0\,a^2 \left(\frac{1}{x}+\frac{1}{y} \right)= -F_0\,a^2 \frac{x+y}{xy}	\,.
\end{equation}
\end{example}

\begin{figure}[h]\begin{center}
\includegraphics[width=0.8\linewidth]{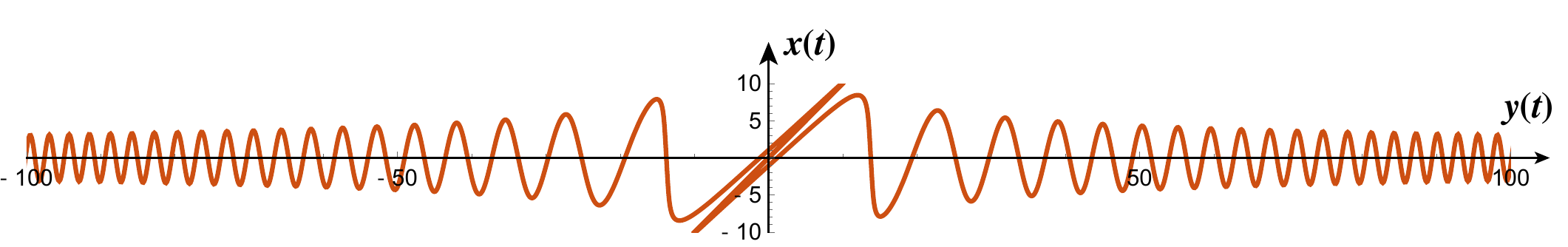}
\caption{Typical solutions of (\ref{Curl-Force-Berry}) with $F/a^3=1$ shown in the $(y,x)$ plane. Asymptotically, $y(t)$ becomes linear in $t$ and $x(t)$ has decaying oscillations to 0. Two solutions are shown here with initial conditions $(x(0)= \pm 10.01, y(0) = 10,\dot x(0)=\dot y(0)=0)$.}
\label{figex1}\end{center}
\end{figure}

The mechanical system with a force given by (\ref{Curl-Force-Berry}) admits different solutions depending on the sign of $F/a^3$. First, we rescale time so that, without without loss of generality the force now reads $\mathbf{F}(x,y)= (xy^2,x^3)$ for $F/a^3<0$, or $\mathbf{F}(x,y)=- (x y^2,x^3)$ for $F/a^3>0$. In the first case, most solutions blow-up in finite time (see \citep{go01} for a general method). In the second case, the solutions can be  periodic for the particular case $x(t)=y(t)\not=0$, for which  the system for both variables is a quartic potential. Otherwise,  the solutions for $y$ tend asymptotically to a linear function of $t$  as $t\to \infty$ and $x(t)$ tends to $0$ as shown in Fig.~\ref{figex1}.

\begin{remark}
Under reasonable regularity assumptions, the first-order linear PDE for $V(x,y)$ obtained in this section always admits local solutions. In particular, since the PDE is of the form $a(x,y) V_{,x} + b(x,y) V_{,y} + c(x,y) V = d(x,y)$ with continuous coefficient functions and $(a(x,y), b(x,y)) \neq (0,0)$, the method of characteristics applies and guarantees local existence of solutions along characteristic curves \citep{Evans1998, John1982, Courant1962}.
\end{remark}

\subsection{Calculation of the generalized potentials for $3$D curl forces} \label{Sec:Darboux-3D}

From \eqref{Curl-Force-3D} we observe that any force field in dimension three can be additively decomposed into a conservative and a non-conservative part:
\begin{equation} \label{Force-c-nc}
    \mathbf{F}(\mathbf{x})= \mathbf{F}_{\text{c}}(\mathbf{x})+\mathbf{F}_{\text{nc}}(\mathbf{x})\,,
    \qquad
    \mathbf{F}_{\text{c}}(\mathbf{x})= -\nabla W(\mathbf{x}) \,,\qquad
    \mathbf{F}_{\text{nc}}(\mathbf{x})=-V(\mathbf{x})\,\nabla U(\mathbf{x}) \,.
\end{equation}
If we are only interested in the non-conservative part of the force field, i.e., if  $\mathbf{F}=\mathbf{F}_{\text{n}}$, the computation of the generalized potentials $V$ and $U$ follows the same procedure as in the two-dimensional case.

\begin{remark}
It should be noted that in the decomposition \eqref{Force-c-nc}, the conservative and non-conservative components are not unique. This is due to the fact that adding a conservative force to a non-conservative force yields a force that is still non-conservative, and subtracting a conservative force from another conservative force results in a force that remains conservative. In other words, $ \mathbf{F}(\mathbf{x})= \tilde{\mathbf{F}}_{\text{c}}(\mathbf{x})+\tilde{\mathbf{F}}_{\text{nc}}(\mathbf{x})$ is an equivalent decomposition if $\tilde{\mathbf{F}}_{\text{nc}}(\mathbf{x})-\mathbf{F}_{\text{nc}}(\mathbf{x})=-(\tilde{\mathbf{F}}_{\text{c}}(\mathbf{x})-\mathbf{F}_{\text{c}}(\mathbf{x}))$ is a conservative force (see Example~\ref{Example-3D}).
\end{remark}

In the decomposition \eqref{Curl-Force-3D}, the three potentials $V$, $U$, and $W$ cannot be uniquely determined. 
Note that $\mathbf{A}=\operatorname{curl}\mathbf{F}=-\nabla V \times \nabla U$. Let us assume that $\nabla V \cdot \nabla U=0$ (the component of $\nabla V$ along $\nabla U$ does not contribute to $\operatorname{curl}\mathbf{F}$ and can be assumed to vanish without affecting the decomposition of the curl force into conservative and non-conservative parts). This orthogonality condition can be interpreted as a gauge choice.
Under the above two assumptions $\{\nabla V,\nabla U,\nabla V \times \nabla U\}$ is a basis for $\mathbb{R}^3$. 
The condition $\nabla V\cdot\mathbf{A}=0$, is a PDE for the unknown scalar field $V$ and can be solved using the method of characteristics. Suppose $\mathbf{x}(s)$ is a curve such that 
\begin{equation} 
	\frac{d \mathbf{x}(s)}{ds} = \mathbf{A}(\mathbf{x}(s))
	\,.
\end{equation}
Along these curves
\begin{equation} 
	\frac{df}{ds} = \nabla f\cdot \frac{d \mathbf{x}(s)}{ds}=\nabla f \cdot \mathbf{A}(\mathbf{x}(s))=0
	\,,
\end{equation}
and hence, $f(\mathbf{x}(s))=\text{constant}$.
Recall that for any three arbitrary vectors $\mathbf{A}$, $\mathbf{B}$, and $\mathbf{C}$:
\begin{equation}
	\mathbf{A} \times (\mathbf{B} \times \mathbf{C}) 
	= (\mathbf{A} \cdot \mathbf{C})\, \mathbf{B} 
	- (\mathbf{A} \cdot \mathbf{B})\, \mathbf{C}\,.
\end{equation}
Thus
\begin{equation} \label{nabla-U}
	\nabla U = \frac{\nabla V \times \operatorname{curl} \mathbf{F}}{\|\nabla V\|^2}\,.
\end{equation}
Therefore
\begin{equation} 
	\mathbf{F}_{\text{c}} = -\nabla W = \mathbf{F} 
	+ \frac{V \nabla V \times \operatorname{curl} \mathbf{F}}{\|\nabla V\|^2}\,.
\end{equation}

\begin{example}\label{Example-3D}
Let us consider the following force field 
\begin{equation} \label{Force-c-nc-Example}
	\mathbf{F}(x, y, z) = -(yz,2xz,xy)
	\,,
\end{equation}
with $\operatorname{curl} \mathbf{F} = (x,0, -z)$.
The condition $\nabla V \cdot \operatorname{curl} \mathbf{F} = 0$ reads
\begin{equation}
	x \, \frac{\partial V}{\partial x}
	- z \, \frac{\partial V}{\partial z} = 0\,.
\end{equation}
This is a linear first-order PDE. The associated characteristic equations are
\begin{equation}
	\frac{dx}{x} = \frac{dz}{-z}, \qquad  y = \text{constant}\,.
\end{equation}
Solving these, one obtains the characteristic curves:
\begin{equation}
	x = C_1 e^s\,, \qquad z = C_2 e^{-s}\,, \qquad y = C_3\,.
\end{equation}
Thus, the invariants (quantities constant along characteristics) are
\begin{equation}
	x z = C_1 C_2 = \text{constant}, \qquad y = \text{constant}\,.
\end{equation}
Therefore, the general solution is
\begin{equation}
	V(x, y, z) = f(x z, y)\,,
\end{equation}
where $f$ is an arbitrary differentiable function.
Now from \eqref{nabla-U} we obtain
\begin{equation}
	\nabla U = \frac{2}{f_2^2 + (x^2 + z^2) f_1^2} 
	\left(- z f_2, (x^2 + z^2) f_1, - x f_2 \right)\,.
\end{equation}
Therefore
\begin{equation} \label{F-c-nc-f}
	\mathbf{F}_{\text{nc}} = \frac{2 f(xz, y)}{f_y^2 + (x^2 + z^2) f_{xz}^2}
	\left(	z f_2, - (x^2 + z^2) f_1, x f_2 \right)\,,
	\qquad
	\mathbf{F}_{\text{c}} = -(yz,2xz,xy)-\mathbf{F}_{\text{nc}}
	\,.
\end{equation}
For the choice $f(xz,y)=y$, we have
\begin{equation}
	\mathbf{F}_{\text{nc}} =  \left( 2yz, 0, 2xy \right)\,,\qquad
	\mathbf{F}_{\text{c}} = (-3yz, -3xz, -3xy)\,.
\end{equation}
For the choice $f(xz,y)=xz$, we have
\begin{equation}
	\tilde{\mathbf{F}}_{\text{nc}} = (0, -2xz,\ 0)\,, \qquad
	\tilde{\mathbf{F}}_{\text{c}} = (-yz, -xz, -xy)\,.
\end{equation}
Note that $\tilde{\mathbf{F}}_{\text{nc}}(\mathbf{x})-\mathbf{F}_{\text{nc}}(\mathbf{x})=-(\tilde{\mathbf{F}}_{\text{c}}(\mathbf{x})-\mathbf{F}_{\text{c}}(\mathbf{x}))= \left( -2yz, -2xz, -2xy \right)$ is a conservative force, and hence, the above two decompositions are equivalent. In other words, both $\mathbf{F}_{\text{c}}$ and $\tilde{\mathbf{F}}_{\text{c}}$ are conservative forces and $\operatorname{curl}\tilde{\mathbf{F}}_{\text{nc}}=\operatorname{curl}\mathbf{F}_{\text{nc}}$. 
In summary, in \eqref{F-c-nc-f} we have a family of equivalent decompositions of the force field \eqref{Force-c-nc-Example} into conservative and non-conservative parts, parametrized by the arbitrary function $f$. 
\end{example}

\begin{remark}[Connection with Helmholtz Decomposition]
It is natural to ask whether the decomposition \eqref{Force-c-nc} is related to the classical Helmholtz decomposition of a vector field into solenoidal (divergence-free) and irrotational (curl-free) components \citep{Helmholtz1858,Helmholtz1867}. 
The Helmholtz decomposition is also related to the Hodge decomposition of differential forms on Riemannian manifolds \citep{Hodge1952,Abraham2012}.
The Helmholtz decomposition, under suitable regularity and decay conditions, states that any sufficiently smooth vector field $\mathbf{F}$ on $\mathbb{R}^3$ that vanishes at infinity can be uniquely written as
\begin{equation}
	\mathbf{F} = -\nabla \phi + \nabla \times \mathbf{A}\,,
\end{equation}
where $\phi$ is a scalar potential and $\mathbf{A}$ is a vector potential. This decomposition separates the field into a curl-free and a divergence-free part.
In contrast, the decomposition \eqref{Force-c-nc} splits a force field into a conservative part $-\nabla W$ and a curl force of the form $-V \nabla U$, where $U$, $V$, and $W$ are generalized scalar potentials. This decomposition is not unique and is defined pointwise rather than through the global elliptic structure that underlies Helmholtz decomposition. 
Moreover, the curl force $-V \nabla U$ is generally neither divergence-free nor orthogonal to the conservative part, and the decomposition is not orthogonal in $L^2$.
Therefore, the decomposition introduced here is conceptually and structurally distinct from the classical Helmholtz decomposition, and it is tailored to the analysis of non-conservative forces arising in mechanics, rather than being a general analytical tool for arbitrary vector fields.
\end{remark}

\subsection{Work performed by curl forces in closed and cyclic motions} \label{Sec:Work}

The \textit{motion of a particle} is a map $\mathbf{x}(t):[t_1,t_2]\to \mathbb{R}^n$ with associated  velocity field $\mathbf{v}(\mathbf{x})=\dot{\mathbf{x}}=\frac{d}{dt}\mathbf{x}(t)$.
The work done by the force field on the particle in this motion is
\begin{equation} 
    W([t_1,t_2])=\int_{t_1}^{t_2} \mathbf{F}(\mathbf{x}(t))\!\cdot\! \dot{\mathbf{x}}(t) dt
    = \int_{\mathbf{x}_1}^{\mathbf{x}_2}  \mathbf{F}(\mathbf{x})\!\cdot\! d\mathbf{x}
    =\int_{\Gamma} \mathbf{F}^\flat=\int_{\Gamma} \boldsymbol{\Omega} \,,
\end{equation}
where $\mathbf{x}_1=\mathbf{x}(t_1)$, $\mathbf{x}_2=\mathbf{x}(t_2)$ and $\Gamma$ is the trajectory of the motion.
A \emph{closed motion} is a motion $\mathbf{x}(t):[t_1,t_2]\to \mathbb{R}^n$ for which $\mathbf{x}(t_1)=\mathbf{x}(t_2)$.
A \emph{cyclic motion} is a motion $\mathbf{x}(t):[t_1,t_2]\to \mathbb{R}^n$ for which $\mathbf{x}(t_1)=\mathbf{x}(t_2)$ and $\dot{\mathbf{x}}(t_1)=\dot{\mathbf{x}}(t_2)$. Clearly, a cyclic motion is closed but the converse is not necessarily true. For a closed motion, $\Gamma\in\mathbb{R}^n$ is a closed curve and encloses a surface $\mathcal{D}\in\mathbb{R}^n$.
Using Stokes' theorem the work done on the particle in a closed motion is calculated as
\begin{equation} 
    W(\Gamma)=\int_{\Gamma} \boldsymbol{\Omega} =\int_{\mathcal{D}}d\boldsymbol{\Omega}
    \,.
\end{equation}
For a curl force for which $d\boldsymbol{\Omega}\neq \mathbf{0}$, the work is non-zero only if $\mathcal{D}$ has non-vanishing area.
Denoting the reverse cyclic motion corresponding to $\Gamma$ by $-\Gamma$, it is straightforward to see that $W(-\Gamma)=-W(\Gamma)$. 
In particular, this implies that curl forces perform zero net work over a cyclic deformation followed by its reverse. This is in contrast with dissipative forces that always perform negative work in such motions.

\subsection{Change of the kinetic energy of a particle under curl forces} \label{Sec:KineticEnergy}

Carath\'eodory \citep{Caratheodory1909} reformulated thermodynamics by expressing the first and second laws in geometric terms on the manifold $\mathcal{M}$ of thermodynamic states. The first law is represented by the vanishing of a differential $1$-form, i.e., $\boldsymbol{\theta}=du-\boldsymbol{\omega}_h-\boldsymbol{\omega}_w=0$, where $u$ is the internal energy, and $\boldsymbol{\omega}_h$ and $\boldsymbol{\omega}_w$ are the heat and work $1$-forms, respectively \citep{Mrugala1978}. An \textit{adiabatic process} is a curve in $\mathcal{M}$ along which $\boldsymbol{\omega}_h$ vanishes. Carath\'eodory’s formulation of the second law asserts that in any neighborhood of a given state, there exist states that are inaccessible via adiabatic processes \citep{Pogliani2000,Frankel2011}. This inaccessibility condition implies that the distribution defined by $\ker \boldsymbol{\omega}_h$ is not completely integrable, thereby constraining the allowable thermodynamic evolutions. Under suitable conditions, Carath\'eodory’s theorem ensures that $\boldsymbol{\omega}_h$ is locally integrable, leading to the existence of an entropy function $s$ and an integrating factor $T$ such that $\boldsymbol{\omega}_h = T\,ds$ \citep{Cooper1967,Boyling1972,Buchdahl2009}.

The kinetic energy of the particle is defined as $K=\frac{1}{2}m\mathbf{v}\cdot \mathbf{v}$, and hence $\frac{d K}{dt}=m \mathbf{v}\cdot \mathbf{a} = \mathbf{F}\cdot \mathbf{v}$.
Thus, $d K =  \mathbf{F}\cdot \mathbf{v}dt =\mathbf{F}\cdot d\mathbf{x} =\mathbf{F}^\flat= \boldsymbol{\Omega}$. This can be rewritten as
\begin{equation} \label{First-Law-C}
    \boldsymbol{\theta} = d K-\boldsymbol{\Omega}=0\,.
\end{equation}
For a particle (or system of particles), the kinetic energy is the internal energy of the system and \eqref{First-Law-C} is a statement of the first law of thermodynamics in the Carath\'eodory's abstract formulation of thermodynamics \citep{Caratheodory1909,Mrugala1978}.
Therefore
\begin{equation} 
	K(t_2)-K(t_1) = \int_{\Gamma} \boldsymbol{\Omega} 
	\,.
\end{equation}
For a $2$D curl force
\begin{equation} 
	K(t_2)-K(t_1) = - \int_{\Gamma} V(\mathbf{x}) \,dU(\mathbf{x}) 
	\,.
\end{equation}
When the generalized potentials $U$ and $V$ are functionally independent, the right-hand side is path dependent and change in the kinetic energy is path dependent as well. Similarly, for a $3$D curl force
\begin{equation} 
	K(t_2)-K(t_1) = - \int_{\Gamma} V(\mathbf{x}) \,dU(\mathbf{x})  - \int_{\Gamma} dW(\mathbf{x})
	\,.
\end{equation}

For a potential force
\begin{equation} 
	K(t_2)-K(t_1) = - \int_{\Gamma} dU(\mathbf{x}) =U(\mathbf{x}_1) - U(\mathbf{x}_2) \,,\quad \text{or}
	\qquad  K(t_2) + U(\mathbf{x}_2) =   K(t_1) +U(\mathbf{x}_1)\,,
\end{equation}
which implies that the total energy $H=K+U$ is conserved.
There is no such energy conservation for curl forces.

\subsection{The local accessibility property of curl forces} \label{Sec:Accessibility}

Let $\boldsymbol{\Omega}\in\Lambda^1(\mathbb{R}^n)$ be a $1$-form. The first-order partial differential equation
\begin{equation}\label{Pfaff}
	\boldsymbol{\Omega}=0\,,
\end{equation}
is known as a \textit{Pfaffian equation}, which is the simplest example of an exterior differential system.  
A $p$-dimensional integral manifold of \eqref{Pfaff} is an immersion (not necessarily an embedding) $f:\mathcal{D}\to\mathbb{R}^n$ such that $f^*\boldsymbol{\Omega}=0$. An integral curve of $\boldsymbol{\Omega}$ is a curve $c:I\to\mathbb{R}^n$, where $I$ is an interval on the real line, satisfying $\boldsymbol{\Omega}(c(t))=0$ for all $t\in I$.  
The Pfaffian equation \eqref{Pfaff} is said to have the \emph{local accessibility} property if, for every point $\mathbf{x}\in\mathbb{R}^n$, there exists a neighborhood $U\subset \mathbb{R}^n$ such that for any point $\mathbf{y}\in U$, there is an integral curve of \eqref{Pfaff} connecting $\mathbf{x}$ and $\mathbf{y}$. Conversely, \eqref{Pfaff} has the \emph{local inaccessibility} property if, in every neighborhood $U$ of any point $\mathbf{x}\in\mathbb{R}^n$, there exists at least one point $\mathbf{y}\in U$ that cannot be reached from $\mathbf{x}$ by any integral curve of \eqref{Pfaff}.

The rank of the Pfaffian equation \eqref{Pfaff} is the integer $r$ such that $\boldsymbol{\Omega}\wedge(d\boldsymbol{\Omega})^r\neq 0$ and $\boldsymbol{\Omega}\wedge(d\boldsymbol{\Omega})^{r+1}=0$. Carath\'eodory's theorem tells us that if the rank of a Pfaffian equation is constant, the Pfaffian equation has the local accessibility property if and only if $r\geq 2$ \citep{Bryant2013}. Therefore, a Pfaffian equation has the local inaccessibility property only when $r=0$ or $r=1$, which correspond to the normal forms $\boldsymbol{\Omega}=d\psi$ and $\boldsymbol{\Omega}=\phi \,d\psi$, respectively. In either case, $\boldsymbol{\Omega}=0$ implies that $\psi=\text{constant}$, which are hypersurfaces---the integral manifolds of maximum possible dimension $n-1$.

\subsubsection{The local in accessibility property of $2$D curl forces} \label{Sec:Accessibility-2D}

Carath\'eodory's theorem tells us that any $2$D curl force with the representation $\mathbf{F}(\mathbf{x})=-V(\mathbf{x})\,\nabla U(\mathbf{x})$ has the inaccessibility property. This has the following physical interpretation. Consider a particle at position $\mathbf{x}(t) \in \mathbb{R}^2$ at time $t$ moving under the influence of a curl force field $\mathbf{F}(\mathbf{x})$. In every neighborhood of $\mathbf{x}(t)$, there exists at least one point $\mathbf{y}$ such that if the particle moves from $\mathbf{x}$ to $\mathbf{y}$ along any path, the curl force performs nonzero work on the particle.

\subsubsection{The local accessibility property of $3$D curl forces} \label{Sec:Accessibility-3D}

Carath\'eodory's theorem states that any $3D$ curl force with the representation $\mathbf{F}(\mathbf{x})=-V(\mathbf{x})\nabla U(\mathbf{x})-\nabla W(\mathbf{x})$ such that none of the three potentials vanishes and  $\mathbf{F}\cdot\operatorname{curl}\mathbf{F}\neq 0$ has the accessibility property. This property has the following physical interpretation. Consider a particle at position $\mathbf{x}(t)\in\mathbb{R}^3$ at time $t$ moving under the influence of the curl force field. There exists a neighborhood $\mathcal{U}$ of $\mathbf{x}(t)$ such that for any $\mathbf{y}\in\mathcal{U}$, there exists a path $\Gamma$ connecting $\mathbf{x}(t)$ to $\mathbf{y}$ such that $\int_{\Gamma} \mathbf{F}(\mathbf{x})\cdot d\mathbf{x}=0$, i.e., the curl force does no work. A $3D$ curl force has the inaccessibility property if it has the representation $\mathbf{F}(\mathbf{x})=-V(\mathbf{x})\nabla U(\mathbf{x})$ or $\mathbf{F}(\mathbf{x})=-\nabla W(\mathbf{x})$, which is a special case of the former.

\begin{remark}
The accessibility condition introduced in this section is equivalent to the classical Frobenius integrability condition applied to the plane field defined by the kernel of the work $1$-form $\boldsymbol{\Omega}$. In particular, the ability to connect points via admissible curves corresponds to the existence of integral manifolds of this distribution, which is guaranteed locally if and only if $\boldsymbol{\Omega} \wedge d\boldsymbol{\Omega} = 0$, or equivalently, if the kernel distribution is involutive under the Lie bracket of vector fields tangent to $\ker \boldsymbol{\Omega}$.\footnote{At each point $p$, the kernel of $\boldsymbol{\Omega}$ is the subspace of tangent vectors annihilated by $\boldsymbol{\Omega}$, i.e., $\ker \boldsymbol{\Omega}_p = \left\{ v \in T_p M \mid \boldsymbol{\Omega}_p(v) = 0 \right\}$. This defines a distribution (i.e., a smoothly varying field of subspaces) on the manifold. A smooth vector field $\mathbf{Y}$ is said to be tangent to $\ker \boldsymbol{\Omega}$ if it takes values in this distribution, that is, $\boldsymbol{\Omega}(\mathbf{Y}(p)) = 0$ for all $p \in M$, or equivalently, $\boldsymbol{\Omega}(\mathbf{Y}) = 0$ everywhere.} Thus, the structure of accessibility is governed by the classical Frobenius theorem on integrability of distributions \cite{Spivak1979, Lee2013, Warner1983}.
\end{remark}

\subsection{The conservative auxiliary force of a curl force} \label{Sec:Hamiltonian}

In this section, we show that any curl force, whether in two or three dimensions, has a corresponding conservative auxiliary force. 
The Hamiltonian associated with this auxiliary force, although not the physical energy, is nevertheless a conserved quantity of the motion.

\subsubsection{Two-dimensional curl forces} \label{Hamiltonian} \label{Sec:Hamiltonian2D}

Any curl force in $2$D has the representation \eqref{Curl-Force-2D}, i.e., $\mathbf{F}(\mathbf{x})=-V(\mathbf{x})\,\nabla U(\mathbf{x})$, where $U$ and $V$ are the generalized potentials of the force field. We are interested in the dynamics of a particle $\mathtt{p}$ under such forces, i.e, the time evolution of its position $\mathbf{x}(t)$, $t\geq 0$. Suppose the initial position and velocity of the particle are given (initial conditions): $\mathbf{x}(0)=\mathbf{x}_0$, $\mathbf{v}(0)=\mathbf{v}_0$. In particular, the initial momentum of the particle is known: $\mathbf{p}(0)=\mathbf{p}_0=m \mathbf{v}_0$, where $m$ is the mass of $\mathtt{p}$.

Let us define an auxiliary conservative force field $\bar{\mathbf{F}}(\mathbf{x})$ by rescaling the original force field with the generalized potential $V$ (assuming that $V(\mathbf{x})\neq 0$):
\begin{equation} 
	\bar{\mathbf{F}}(\mathbf{x}) =\frac{1}{V(\mathbf{x})} \mathbf{F}(\mathbf{x})\,,\qquad 
	 \bar{\mathbf{F}}(\mathbf{x})=-\nabla U(\mathbf{x})\,.
\end{equation}
The particle $\mathtt{p}$, in general, would have a different dynamics under $\bar{\mathbf{F}}$, which we call the auxiliary dynamics. Let us denote its trajectory by $\bar{\mathbf{x}}(t)$, $t\geq 0$. The velocity field corresponding to the rescaled force field is denoted by $\bar{\mathbf{v}}(t)$ and its momentum by $\bar{\mathbf{p}}=m\bar{\mathbf{v}}$. 
The initial conditions are independent of the force field, and hence, $\bar{\mathbf{x}}(0)=\mathbf{x}_0$, $\bar{\mathbf{v}}(0)=\mathbf{v}_0$. Also, the initial momentum is independent of the force field as well: $\bar{\mathbf{p}}(0)=\mathbf{p}_0=m \mathbf{v}_0$.
One has the following Hamiltonian corresponding to the conservative auxiliary force field:
\begin{equation} \label{Hamiltonian-Auxiliary}
	H(\bar{\mathbf{x}},\bar{\mathbf{p}})
	= \frac{\bar{\mathbf{p}}\cdot\bar{\mathbf{p}}}{2m}+U(\bar{\mathbf{x}})    \,.
\end{equation}
We call $H$ the auxiliary Hamiltonian. Hamilton's equations for the auxiliary motion read 
\begin{equation} \label{Hamilton-Equations-Auxiliary}
\begin{dcases}
	-\frac{\partial H}{\partial \bar{\mathbf{x}}}
	=\dot{\bar{\mathbf{p}}}
	=-\nabla U(\bar{\mathbf{x}}) = \bar{\mathbf{F}}(\bar{\mathbf{x}})\,, \\
	\frac{\partial H}{\partial \bar{\mathbf{p}}}=\frac{\bar{\mathbf{p}}}{m}=\bar{\mathbf{v}}\,.
\end{dcases}
\end{equation}
Note that $\frac{dH}{dt}=0$, and hence $H$ is a constant of motion, i.e.,
\begin{equation} 
	\frac{\bar{\mathbf{p}}(t)\cdot\bar{\mathbf{p}}(t)}{2m}+U(\bar{\mathbf{x}}(t))
	=\frac{1}{2}m\, \bar{\mathbf{v}}(t)\cdot\bar{\mathbf{v}}(t)+U(\bar{\mathbf{x}}(t))    
	=\frac{1}{2}m \mathbf{v}_0\cdot\mathbf{v}_0 +U(\mathbf{x}_0)
	\,.
\end{equation}

From $\bar{\mathbf{F}}=m\dot{\bar{\mathbf{v}}}$, and knowing that for the original force field $\mathbf{F}=m\dot{\mathbf{v}}=\dot{\mathbf{p}}$ (Newton's law of motion is independent of whether a Hamiltonian exists), one concludes that
\begin{equation} \label{Momentum}
	\dot{\bar{\mathbf{p}}}=\frac{1}{V(\mathbf{x})}\dot{\mathbf{p}}\,.
\end{equation}
Thus
\begin{equation} 
	\bar{\mathbf{p}}(t) = \mathbf{p}_0
	+\int_0^t \frac{\dot{\mathbf{p}}(\xi)}{V(\mathbf{x}(\xi))}\,d\xi
	= \mathbf{p}_0
	-\int_0^t  \nabla U(\mathbf{x}(\xi)) \,d\xi
	\,,
\end{equation}
where $\mathbf{p}_0=\bar{\mathbf{p}}(0)$ and the equation of motion $\dot{\mathbf{p}}=-V \nabla U$ was used in the second equality.
Recalling that $\bar{\mathbf{p}}(t) =m \bar{\mathbf{v}}(t) = m\dot{\bar{\mathbf{x}}}(t)$, we have
\begin{equation} \label{Auxiliary-Motion}
	\bar{\mathbf{x}}(t) = \mathbf{x}_0+\mathbf{v}_0 \,t
	-\frac{1}{m}\int_0^t \int_0^\tau  \nabla U(\mathbf{x}(\xi)) \,d\xi\,d\tau
	\,.
\end{equation}
Therefore, the auxiliary Hamiltonian \eqref{Hamiltonian-Auxiliary} is simplified to read
\begin{equation}  \label{Hamiltonian-2D}
\begin{aligned}
	H & =  \frac{1}{2m} \left[ |\mathbf{p}_0|^2
	-2 \mathbf{p}_0\cdot \int_0^t \nabla U(\mathbf{x}(\xi)) \,d\xi
	+\int_0^t\int_0^t 
	\nabla U(\mathbf{x}(\xi)) \cdot \nabla U(\mathbf{x}(\eta))\,d\xi\,d\eta \right] \\
	& \quad +U\!\left( \mathbf{x}_0+t \mathbf{v}_0
	-\frac{1}{m}\int_0^t \int_0^\tau  \nabla U(\mathbf{x}(\xi)) \,d\xi\,d\tau \right)    \,.
\end{aligned}
\end{equation}
Note that $H$ is a conserved quantity of motion under the curl force, although it is not the physical energy. Moreover, due to the fact that it is defined implicitly through integral equations, this conserved quantity is not like the traditional constants of motion that appears in integrability theory. Its existence does not preclude chaos and cannot be used to reduce the dynamics to manifolds of lower dimensions on which this quantity is conserved.

\begin{example}
Let us consider the force field \eqref{Curl-Force-Berry} of Example~\ref{Force-Field-Calculation}.
In this example, the rescaled force field has the following form
\begin{equation} 
	\bar{\mathbf{F}}(\mathbf{x})
	=\frac{1}{V(\mathbf{x})} \mathbf{F}(\mathbf{x})
	= -F_0a^2\left(\frac{1}{x^2},\frac{1}{y^2}\right)	\,.
\end{equation}
The auxiliary motion \eqref{Auxiliary-Motion} is given as
\begin{equation} 
	\bar{\mathbf{x}}(t) = \mathbf{x}_0+ \mathbf{v}_0 \,t
	-\frac{F_0a^2}{m}\int_0^t \int_0^\tau  \left( \frac{1}{x^2(\xi)}\,,\frac{1}{y^2(\xi)}\right) \,d\xi\,d\tau
	\,.
\end{equation}
The Hamiltonian \eqref{Hamiltonian-2D} is simplified to read
\begin{equation} 
\begin{aligned}
	H & =  \frac{1}{2m} \Bigg\{ p_{x0}^2+p_{y0}^2
	+2   \int_0^t \left[\frac{p_{x0}}{x^2(\xi)}+\frac{p_{y0}}{y^2(\xi)}\right] d\xi 
	- F_0^2 a^4 \int_0^t\int_0^t  \left[\frac{1}{x^2(\xi) \,x^2(\eta)}
	+\frac{1}{y^2(\xi) \,y^2(\eta)}\right] d\xi\,d\eta
	\Bigg \} \\
	& \quad + U\!\left( \mathbf{x}_0+ \mathbf{v}_0 \,t
	+\frac{F_0a^2}{m}\int_0^t \int_0^\tau  
	\left(\frac{1}{x^2(\xi)}\,,\frac{1}{y^2(\xi)}\right) \,d\xi\,d\tau \right) 
	  \,,
\end{aligned}
\end{equation}
where $\mathbf{p}_0=(p_{x0},p_{y0})$ and $U$ is given in \eqref{U-Example}.
$H$ is a conserved quantity for the motion under the curl force \eqref{Curl-Force-Berry}.
\end{example}

\subsubsection{Three-dimensional curl forces} \label{Sec:Hamiltonian3D}

Any curl force in $3$D has the representation \eqref{Curl-Force-3D}, and hence, the rescaled force field 
\begin{equation} 
	\bar{\mathbf{F}}(\mathbf{x})
	=\frac{\mathbf{F}(\mathbf{x})+\nabla W(\mathbf{x})}{V(\mathbf{x})} \,,
\end{equation}
is conservative, i.e., $ \bar{\mathbf{F}}(\mathbf{x})=-\nabla U(\mathbf{x})$. Let us denote the velocity field corresponding to the auxiliary force field by $\bar{\mathbf{v}}(\mathbf{x})$ and its momentum by $\bar{\mathbf{p}}=m\bar{\mathbf{v}}$. One still has the Hamiltonian \eqref{Hamiltonian-Auxiliary} corresponding to the auxiliary force.
Hamilton's equations are given in \eqref{Hamilton-Equations-Auxiliary}. 
The relations between the original and auxiliary dynamics are still described by Eqs.~\eqref{Momentum}-\eqref{Auxiliary-Motion}.
The auxiliary motion still has the form \eqref{Auxiliary-Motion}. Therefore, the auxiliary Hamiltonian \eqref{Hamiltonian-Auxiliary} simplifies to read:
\begin{equation} 
\begin{aligned}
	H(\mathbf{x},\mathbf{p})
	& = \frac{1}{2m} \left[ \mathbf{p}_0\cdot \mathbf{p}_0
	+2\mathbf{p}_0\cdot \int_0^t \frac{\dot{\mathbf{p}}(\xi)}{V(\mathbf{x}(\xi))}\,d\xi
	+\int_0^t \frac{\dot{\mathbf{p}}(\xi)}{V(\mathbf{x}(\xi))}\,d\xi
	\cdot \int_0^t \frac{\dot{\mathbf{p}}(\eta)}{V(\mathbf{x}(\eta))}\,d\eta \right]
	+U(\bar{\mathbf{x}}) \\
	& = \frac{1}{2m}  \left[ |\mathbf{p}_0|^2
	+2\mathbf{p}_0\cdot \int_0^t \frac{\dot{\mathbf{p}}(\xi)}{V(\mathbf{x}(\xi))}\,d\xi
	+\int_0^t\int_0^t 
	\frac{\dot{\mathbf{p}}(\xi)\cdot\dot{\mathbf{p}}(\eta)}{V(\mathbf{x}(\xi))\,V(\mathbf{x}(\eta))}
	\,d\xi\,d\eta
	\right]
	+U(\bar{\mathbf{x}})    \,.
\end{aligned}
\end{equation}
Using the equations of motion $\dot{\mathbf{p}}=-V \nabla U-\nabla W$, Hamiltonian is simplified to read
\begin{equation} 
\begin{aligned}
	H & =  \frac{1}{2m} \Bigg\{ |\mathbf{p}_0|^2 -2 \mathbf{p}_0\cdot 
	\int_0^t \left[\nabla U(\mathbf{x}(\xi))+\frac{\nabla W(\mathbf{x}(\xi))}{V(\mathbf{x}(\xi))} \right] 
	\,d\xi \\
	&\qquad +\int_0^t\int_0^t 
	\left[\nabla U(\mathbf{x}(\xi))+\frac{\nabla W(\mathbf{x}(\xi))}{V(\mathbf{x}(\xi))} \right]
	\!\cdot\! \left[\nabla U(\mathbf{x}(\eta))+\frac{\nabla W(\mathbf{x}(\eta))}{V(\mathbf{x}(\eta))} 
	\right]
	\,d\xi\,d\eta \Bigg\}\\
	&\qquad + U\!\left( \mathbf{x}_0+t \mathbf{v}_0
	-\frac{1}{m}\int_0^t \int_0^\tau  \nabla U(\mathbf{x}(\xi)) \,d\xi\,d\tau \right)   \,.
\end{aligned}
\end{equation}
This is a conserved quantity of motion under an arbitrary three-dimensional curl force, although it does not represent the physical energy.

\section{Conclusions}  \label{Sec:Conclusions}

In this paper, we investigated the dynamics of particles under curl forces, a class of non-conservative, non-dissipative forces. 
We first pointed out that the fundamental quantity associated with a curl force is its work $1$-form, which turns out to be the flat force field, that is, the $1$-form representation (or proxy) of the force vector field.
Using the Darboux classification of differential $1$-forms, we established that any two-dimensional curl force has at most two generalized potentials, while in three dimensions, it has at most three. This classification provides a systematic framework for understanding the structure of curl forces and their associated work $1$-forms.
It was shown that one feature distinguishing curl forces from potential forces is their ability to perform non-zero work over certain classes of closed paths.
We also discussed the accessibility property of curl forces in dimensions two and three.
Furthermore, we demonstrated that for any given curl force field, a corresponding conservative force field can be constructed. This conservative auxiliary force enabled the definition of an auxiliary Hamiltonian for curl forces, which is a nonlocal functional of motion and is a conserved quantity of motion. Our formulation offers a new perspective on the representation of curl forces in dimensions two and three and their underlying mathematical properties.


\section*{Acknowledgments}

This work originated from a chance encounter with Michael Berry at the Venice airport, during which a discussion unfolded that suggested a potential connection between curl forces and Cauchy elasticity. We are grateful to Michael Berry for making the original connection between Cauchy elasticity and curl forces. We also benefitted from a discussion with Francesco Fedele.

\bibliographystyle{abbrvnat}

\begin{thebibliography}{37}
\providecommand{\natexlab}[1]{#1}
\providecommand{\url}[1]{\texttt{#1}}
\expandafter\ifx\csname urlstyle\endcsname\relax
  \providecommand{\doi}[1]{doi: #1}\else
  \providecommand{\doi}{doi: \begingroup \urlstyle{rm}\Url}\fi

\bibitem[Abraham et~al.(2012)Abraham, Marsden, and Ratiu]{Abraham2012}
R.~Abraham, J.~E. Marsden, and T.~Ratiu.
\newblock \emph{Manifolds, Tensor Analysis, and Applications}, volume~75 of
  \emph{Applied Mathematical Sciences}.
\newblock Springer Science \& Business Media, 2012.

\bibitem[Arnold and Khesin(1998)]{ArnoldKhesin1998}
V.~I. Arnold and B.~A. Khesin.
\newblock \emph{Topological Methods in Hydrodynamics}, volume 125 of
  \emph{Applied Mathematical Sciences}.
\newblock Springer, 1998.

\bibitem[Berry and Shukla(2012)]{Berry2012}
M.~V. Berry and P.~Shukla.
\newblock Classical dynamics with curl forces, and motion driven by
  time-dependent flux.
\newblock \emph{Journal of Physics A}, 45\penalty0 (30):\penalty0 305201, 2012.

\bibitem[Berry and Shukla(2013)]{Berry2013}
M.~V. Berry and P.~Shukla.
\newblock Physical curl forces: dipole dynamics near optical vortices.
\newblock \emph{Journal of Physics A: Mathematical and Theoretical},
  46\penalty0 (42):\penalty0 422001, 2013.

\bibitem[Berry and Shukla(2015)]{Berry2015}
M.~V. Berry and P.~Shukla.
\newblock Hamiltonian curl forces.
\newblock \emph{Proceedings of the Royal Society A: Mathematical, Physical and
  Engineering Sciences}, 471\penalty0 (2176):\penalty0 20150002, 2015.

\bibitem[Boyling(1972)]{Boyling1972}
J.~Boyling.
\newblock An axiomatic approach to classical thermodynamics.
\newblock \emph{Proceedings of the Royal Society of London. A}, 329\penalty0
  (1576):\penalty0 35--70, 1972.

\bibitem[Bryant et~al.(2013)Bryant, Chern, Gardner, Goldschmidt, and
  Griffiths]{Bryant2013}
R.~L. Bryant, S.-S. Chern, R.~B. Gardner, H.~L. Goldschmidt, and P.~A.
  Griffiths.
\newblock \emph{Exterior Differential Systems}, volume~18.
\newblock Springer Science \& Business Media, 2013.

\bibitem[Buchdahl(2009)]{Buchdahl2009}
H.~A. Buchdahl.
\newblock \emph{The Concepts of Classical Thermodynamics}.
\newblock Cambridge University Press, 2009.

\bibitem[Cantarella et~al.(2002)Cantarella, DeTurck, and
  Gluck]{CantarellaDeturckGluck2002}
J.~Cantarella, D.~DeTurck, and H.~Gluck.
\newblock Vector calculus and the topology of domains in 3-space.
\newblock \emph{American Mathematical Monthly}, 109\penalty0 (5):\penalty0
  409--442, 2002.
\newblock \doi{10.1080/00029890.2002.11919870}.

\bibitem[Carath{\'e}odory(1909)]{Caratheodory1909}
C.~Carath{\'e}odory.
\newblock Untersuchungen {\"u}ber die {G}rundlagen der {T}hermodynamik.
\newblock \emph{Mathematische Annalen}, 67\penalty0 (3):\penalty0 355--386,
  1909.

\bibitem[Cooper(1967)]{Cooper1967}
J.~L. Cooper.
\newblock The foundations of thermodynamics.
\newblock \emph{Journal of Mathematical Analysis and Applications}, 17\penalty0
  (1):\penalty0 172--193, 1967.

\bibitem[Courant and Hilbert(1962)]{Courant1962}
R.~Courant and D.~Hilbert.
\newblock \emph{Methods of Mathematical Physics. Volume II: Partial
  Differential Equations}.
\newblock Interscience Publishers, New York, 1962.

\bibitem[Darboux(1882)]{Darboux1882}
G.~Darboux.
\newblock Sur le probleme de {P}faff.
\newblock \emph{Bulletin des sciences math{\'e}matiques et astronomiques},
  6\penalty0 (1):\penalty0 14--36, 1882.

\bibitem[Duff(1956)]{Duff1956}
G.~F.~D. Duff.
\newblock \emph{Partial Differential Equations}.
\newblock University of Toronto Press, 1956.

\bibitem[Evans(1998)]{Evans1998}
L.~C. Evans.
\newblock \emph{Partial Differential Equations}, volume~19 of \emph{Graduate
  Studies in Mathematics}.
\newblock American Mathematical Society, Providence, RI, 1998.

\bibitem[Frankel(2011)]{Frankel2011}
T.~Frankel.
\newblock \emph{The Geometry of Physics: An Introduction}.
\newblock Cambridge University Press, 2011.

\bibitem[Goriely(2001)]{go01}
A.~Goriely.
\newblock \emph{Integrability and Nonintegrability of Dynamical Systems}.
\newblock World Scientific Publishing Company, 2001.

\bibitem[Helmholtz(1858)]{Helmholtz1858}
H.~Helmholtz.
\newblock {\"U}ber integrale der hydrodynamischen gleichungen, welche den
  wirbelbewegungen entsprechen.
\newblock \emph{Journal f{\"u}r die reine und angewandte Mathematik},
  55:\penalty0 25--55, Jan. 1858.

\bibitem[Helmholtz(1867)]{Helmholtz1867}
H.~Helmholtz.
\newblock On integrals of the hydrodynamical equations, which express
  vortex-motion.
\newblock \emph{Philosophical Magazine and Journal of Science}, 33\penalty0
  (226):\penalty0 485--512, 1867.

\bibitem[Hodge(1952)]{Hodge1952}
W.~V.~D. Hodge.
\newblock \emph{The Theory and Applications of Harmonic Integrals}.
\newblock Cambridge University Press, Cambridge, 1952.

\bibitem[John(1982)]{John1982}
F.~John.
\newblock \emph{Partial Differential Equations}, volume~1 of \emph{Applied
  Mathematical Sciences}.
\newblock Springer, New York, 1982.

\bibitem[Kirillov(2021)]{Kirillov2021}
O.~N. Kirillov.
\newblock \emph{Nonconservative Stability Problems of Modern Physics},
  volume~14.
\newblock Walter de Gruyter GmbH \& Co KG, 2021.

\bibitem[Lee(2013)]{Lee2013}
J.~M. Lee.
\newblock \emph{Introduction to Smooth Manifolds}, volume 218 of \emph{Graduate
  Texts in Mathematics}.
\newblock Springer, 2nd edition, 2013.

\bibitem[Moffatt(1969)]{Moffatt1969}
H.~K. Moffatt.
\newblock The degree of knottedness of tangled vortex lines.
\newblock \emph{Journal of Fluid Mechanics}, 35\penalty0 (1):\penalty0
  117--129, 1969.
\newblock \doi{10.1017/S0022112069000991}.

\bibitem[Moffatt and Tsinober(1992)]{MoffattTsinober1992}
H.~K. Moffatt and A.~Tsinober.
\newblock Helicity in laminar and turbulent flow.
\newblock \emph{Annual Review of Fluid Mechanics}, 24:\penalty0 281--312, 1992.
\newblock \doi{10.1146/annurev.fl.24.010192.001433}.

\bibitem[Mrugala(1978)]{Mrugala1978}
R.~Mrugala.
\newblock Geometrical formulation of equilibrium phenomenological
  thermodynamics.
\newblock \emph{Reports on Mathematical Physics}, 14\penalty0 (3):\penalty0
  419--427, 1978.

\bibitem[Pogliani and Berberan-Santos(2000)]{Pogliani2000}
L.~Pogliani and M.~N. Berberan-Santos.
\newblock Constantin {C}arath{\'e}odory and the axiomatic thermodynamics.
\newblock \emph{Journal of Mathematical Chemistry}, 28:\penalty0 313--324,
  2000.

\bibitem[Slebodzinski(1970)]{slebodzinski1970exterior}
W.~Slebodzinski.
\newblock Exterior forms and their applications.
\newblock \emph{Mathematical Monographs, 52, Mathematical Institute of Polish
  Academy of Sciences}, 1970.

\bibitem[Spivak(1979)]{Spivak1979}
M.~Spivak.
\newblock \emph{A Comprehensive Introduction to Differential Geometry, Vol. 1}.
\newblock Publish or Perish, 2nd edition, 1979.

\bibitem[Sternberg(1999)]{Sternberg1999}
S.~Sternberg.
\newblock \emph{Lectures on Differential Geometry}, volume 316.
\newblock American Mathematical Soc., 1999.

\bibitem[Suhubi(2013)]{Suhubi2013}
E.~Suhubi.
\newblock \emph{Exterior Analysis: Using Applications of Differential Forms}.
\newblock Elsevier, 2013.

\bibitem[Trueba and nada(1996)]{TruebaRanada1996}
J.~L. Trueba and A.~F.~R. nada.
\newblock The electromagnetic helicity.
\newblock \emph{European Journal of Physics}, 17\penalty0 (3):\penalty0
  141--144, 1996.
\newblock \doi{10.1088/0143-0807/17/3/005}.

\bibitem[Warner(1983)]{Warner1983}
F.~W. Warner.
\newblock \emph{Foundations of Differentiable Manifolds and Lie Groups},
  volume~94 of \emph{Graduate Texts in Mathematics}.
\newblock Springer, 1983.

\bibitem[Yavari(2013)]{Yavari2013}
A.~Yavari.
\newblock Compatibility equations of nonlinear elasticity for
  non-simply-connected bodies.
\newblock \emph{Archive for Rational Mechanics and Analysis}, 209\penalty0
  (1):\penalty0 237--253, 2013.

\bibitem[Yavari and Goriely(2024)]{yavari2024nonlinear}
A.~Yavari and A.~Goriely.
\newblock Nonlinear cauchy elasticity.
\newblock \emph{arXiv preprint arXiv:2412.17090}, 2024.

\bibitem[Ziegler(1977)]{Ziegler1977}
H.~Ziegler.
\newblock \emph{Principles of Structural Stability}.
\newblock Springer, Basel, 1977.

\end{thebibliography}

\end{document}